\documentclass[floats,floatfix,showpacs,amssymb,prl,twocolumn,superscriptaddress,nofootinbib]{revtex4-2}

\usepackage{graphicx}
\usepackage{dcolumn}
\usepackage{bm}
\usepackage{amsmath}
\usepackage[colorlinks=true,linkcolor=blue,citecolor=blue,urlcolor=blue]{hyperref}
\usepackage{dsfont}

\usepackage{xcolor}
\usepackage[normalem]{ulem}

\newcommand{\bfr}{{\bf r}}

\newcommand{\bfv}{{\bf v}}

\newcommand\comment[1]{}
\newcommand{\vrel}{{v_{\rm rel}}}

\begin{document}


\title{Numerical evolution of self-gravitating halos of self-interacting dark matter}

\author{Marc Kamionkowski}
\email{kamion@jhu.edu}
\affiliation{William H.\ Miller III Department of Physics \& Astronomy, Johns Hopkins University, 3400 N.\ Charles St., Baltimore, MD 21218, USA}

\author{Kris Sigurdson}
\email{krs@phas.ubc.ca}
\affiliation{Department of Physics and Astronomy, University of British Columbia, Vancouver, BC V6T 1Z1, Canada}

\author{Oren Slone}
\email{orenslone@tauex.tau.ac.il}
\affiliation{Raymond and Beverly Sackler School of Physics and Astronomy, Tel Aviv University, Tel-Aviv 69978, Israel}

\setcounter{footnote}{0}
\def\thefootnote{\arabic{footnote}}

\begin{abstract}
We discuss a modification of a recently developed numerical scheme for evolving spherically symmetric self-gravitating systems to include the effects of self-interacting dark matter.  The approach is far more efficient than traditional N-body simulations and cross sections with different dependencies on velocity and scattering-angle are easily accommodated.  To demonstrate, we provide results of a simulation, which runs quickly on a personal computer, that shows the expected initial flattening of the inner region of an NFW halo as well as the later gravothermal collapse instability that leads to a dense core at the galactic center.  We note that this approach can also be used, with some augmentation, to simulate the dynamics in globular clusters by modeling gravitational hard scattering as a self-interaction.
\end{abstract}

\maketitle


Self-interacting dark matter (SIDM) \cite{Spergel:1999mh,Tulin:2017ara,Buckley:2017ijx} has long been considered as a possible explanation for discrepancies between observed properties of dark-matter halos---particularly those for dwarf galaxies \cite{Kaplinghat:2015aga,Kamada:2016euw,Zentner:2022xux,Correa:2020qam,Zeng:2021ldo,Correa:2022dey,Nadler:2023nrd,Yang:2022mxl}---and those expected if dark matter is collisionless.  

The dynamics of self-gravitating systems of SIDM are interesting.  If the probability for a particle with a trajectory that passes near the center to interact is larger than unity over the lifetime of the halo, then SIDM transports heat between the inner and outer regions of the halo. Initially, this heat transfer tends to smooth out velocity-dispersion and density gradients throughout the halo's central region~\cite{Colin:2002nk}. If the halo further evolves over many scattering timescales, heat flow out of the center drives the system towards the runaway process of gravothermal core-collapse during which the central density grows rapidly~\cite{Kochanek:2000pi,Balberg:2001qg}. The dynamics of this process is often approximated by a fluid approach whereby moments of the Boltzmann equation are solved for properties of the system such as density and temperature. However, the validity of a fluid approximation can be called into question, as the particles may still travel distances between scatters that are much larger than the typical scales of the system (unless the self-interaction cross section is huge).  Additionally, in a typical halo the mean-free path may be long in the outskirts but short near the center. Therefore, a full description of the dynamics can become complicated \cite{Dave:2000ar,Koda:2011yb,Essig:2018pzq,Nishikawa:2019lsc,Slone:2021nqd,Tulin:2017ara,Outmezguine:2022bhq,Yang:2022zkd,Gad-Nasr:2023gvf}, and parameters in analytic models are typically fit to N-body simulations \cite{Kummer:2019yrb,Fischer:2020uxh,Fischer:2024eaz,Mace:2024uze,Palubski:2024ibb,Mace:2025fuz}. These simulations require significant computational resources, and the self-interactions introduce several new technical difficulties, in addition to those common to all N-body simulations. These arise, for example, from issues with energy conservation and with sampling the local phase-space distributions when evaluating collision kernels.

Here we show how a recently developed approach to N-body dynamics of spherical systems \cite{Kamionkowski:2025uae} can be augmented to include self-interactions. The new approach capitalizes on the fact that a spherical system can be described by a three-dimensional phase space, rather than the six-dimensional phase space evolved in traditional N-body simulations \cite{Kamionkowski:2025uae}. The algorithm allows the phase space for a system composed of $10^6$ particles to be evolved on a personal computer in roughly a minute timescale per dynamical time. It moreover sidesteps or greatly diminishes many challenges (e.g., force-softening, energy conservation, phase-space sampling) in traditional N-body simulations. We illustrate with some initial results and convergence tests. Given the vast parameter space for SIDM models, as well as the large mass range for astrophysical dark-matter halos, we leave the full potential of this new tool to be explored in future work.\footnote{Our work bears some resemblance to, but differs in detail from, the approach described recently in Ref.~\protect\cite{Gurian:2025zpc}.}

Below we first reprise the algorithm of Ref.~\cite{Kamionkowski:2025uae}.  We then show how to introduce self-interactions into the calculations and show results and convergence tests before closing with concluding remarks.

\textit{Initial Conditions.} Following Ref.~\cite{Kamionkowski:2025uae},
we work with symmetric self-gravitating systems of density $\rho(r)$  with $\rho(r)=0$ at radii $r>R$ and total mass $M_h=M(R) \equiv 4\pi\int_0^R\, dr\,r^2\,\rho(r)$.  The relative gravitational potential is $\Psi(r) = -\Phi(r) + \Phi_0$, with $\Phi(r)$ the usual gravitational potential, so that the Poisson equation is $\nabla^2\Psi(r)= -4\pi G\rho(r)$.  We impose boundary conditions $\Psi(r) \to 0$ as $r\to \infty$.  A particle of velocity $v$ has a relative energy ${\cal E} = \Psi(r)-(1/2)v^2$.  The most loosely bound orbits have ${\cal E} \to 0^+$, and the most tightly bound have the largest ${\cal E}$.

The system is initialized by populating a halo with $N$ simulation particles drawn from an initial phase-space density,
\begin{equation}
     f({\bf x},{\bf v}) = f_{\rm init}({\cal E}) = \frac{1}{m_p\sqrt{8}\pi^2} \frac{d}{d{\cal E}}\int_0^{\cal E}\, \frac{d\Psi}{\sqrt{{\cal E}-\Psi}} \frac{d\rho}{d\Psi},
\end{equation}
where $\rho(r)$ is the initial density profile.  Here, the phase-space density is defined so that the mass density at some position ${\bf x}$ is $\rho({\bf x}) = m_p \int\,  d^3 v f\left(\Psi(r)-v^2/2\right)$.  In the real halo, $m_p$ would be the particle mass, but in the following we will take it to be the mass $m_p=M_h/N$ within a radial particle shell in the simulation.  The radius $r_i$ of the $i$th particle is drawn from a radial distribution proportional to $r^2 \rho(r)$.  The magnitude $v_i$ of its velocity is then drawn from a velocity distribution proportional to $v^2 f\left(\Psi(r)-v^2/2\right)$. The angle $\theta_i$ the velocity makes with the radial direction is drawn from a uniform distribution in $-1<\cos\theta_i<1$, where $\theta$ is the angle the velocity makes with the radial direction. The choice of a uniform distribution in $\cos\theta$ supposes that the initial phase-space density depends only on the particle energy, but a nontrivial dependence on angular momentum can be easily incorporated.

{\it Dynamics.}  With spherical symmetry, the distance $r_i$ of the $i$th particle from the origin is governed by
\begin{equation}
     \ddot r_i = -\frac{GM(<r_i)}{r_i^2} + \frac{\ell_i^2}{r_i^3},
\label{eqn:eom}
\end{equation}
where $\ell_i$ is the angular momentum per unit mass of the $i$th particle, and the dot denotes derivative with respsect to time.  In each time step in the simulation, the particles are first ordered by radius so that the mass $M(<r_i)=m_p (i-1)$ enclosed within a radius $r_i$ is simply the number of particles at smaller radii times the particle mass.  Each particle is then advanced according to Eq.~(\ref{eqn:eom}).  We then iterate the reordering by radius and continue.  In the absence of self-interactions, our code (available at {\tt github.com/kris-sigurdon/NSphere}) preserves the density profile of a self-consistent self-gravitating halo for hundreds of dynamical times, and particle energies are preserved fairly precisely on the same timescales.

{\it Scattering.}  Self-interactions imply a nonzero differential cross section
$(d\sigma/d\alpha)$ which is a function of the relative velocity $\vrel$ between scattering particles and the outgoing scattering angle $\alpha$. Here we restrict our attention to elastic scattering and identical particles, but the generalizations to inelastic interactions \cite{Tucker-Smith:2001myb,Krnjaic:2014xza,Vogelsberger:2018bok,Chua:2020svq,ONeil:2022szc,Roy:2023zar} or different particle types are both straightforward.  

In the simulation, the probability that any given particle , which we call particle ``1,'' at radius $r_1$ and velocity $\bfv_1$, interacts in a small time interval $\Delta t$ with another particle is
\begin{equation}
     P_1(r_1,v_1,c_{\theta_1}) =\frac{\Delta t}{2} \int d^3v_i f(r_i,v_i,c_{\theta_i}) \sigma_{\rm tot}( \vrel_{,i})\frac{m_p}{m}\vrel_{,i},
\label{eqn:prob}     
\end{equation}
where $\vrel_{,i}=|\bfv_1-\bfv_i|$ is the relative velocity between particle 1 and the other particle (the $i$th particle) it interacts with, and $\sigma_{\rm tot}(\vrel) = \int_{-1}^1\, d\cos\alpha (d\sigma/d\cos\alpha)$ is the total cross section. The factor $(m_p/m)$ (where $m$ is the elementary-particle mass) in Eq.~(\ref{eqn:prob}) accounts for the re-scaling of the cross section for simulation particles relative to that ($\sigma_{\rm tot}$) for dark-matter particles.  

Although the evolution of a collisionless system does not depend on the azimuthal angle $\phi_i$ of the particle velocity around the radial direction, the kinematics of the 2-body interaction depends on the relative azimuthal angle.  We therefore assign to each particle a random $\phi_i$ drawn uniformly from $0$ to $2\pi$.  The velocities can be written explicitly as $\bfv_1 = v_1(s_{\theta_1},0,c_{\theta_1})$ and $\bfv_i=  v_i(s_{\theta_i} c_{\phi}, s_{\theta_i} s_\phi, c_{\theta_i})$, with the shorthands $s_\xi=\sin\xi$ and $c_\xi=\cos\xi$.  The phase-space density for the discretized system is thus
\begin{eqnarray}
    f(r,v,c_\theta,\phi) &=& \frac{1}{4\pi} \sum_i \frac{1}{v_i^2} \frac{1}{r_i^2} \delta_D(r-r_i) \delta_D(v-v_i) \nonumber \\
    & & \times \delta_D(c_\theta-c_{\theta_i}) \delta_D(\phi-\phi_i),
\label{eqn:discrete}    
\end{eqnarray}
where $\delta_D(x)$ is the Dirac delta function.  Eq.~(\ref{eqn:discrete}) can be verified by noting that when integrated the total halo mass $M_h = m_p \int d^3 v \int d^3r f(\bfr,\bfv)$ is recovered.

Strictly speaking, the phase-space density at any given point can never be determined with a realization with any finite number of particles; it can only be determined after it has been coarse-grained over some small volume $\Delta V$.  With spherical symmetry imposed, as we do here, the phase-space density at the position $r_1$ can be approximated by coarse graining over some range $\Delta r$ of radii.  We implement this here with a top-hat kernel of width $\Delta r$ (although other kernels can also be tried).  More precisely, we replace the discretized phase-space density in Eq.~(\ref{eqn:discrete}) by $(\Delta r)^{-1}\int_r^{r+\Delta r} \, dr\, f(r,v,c_\theta,\phi)$.\footnote{We choose this one-sided interval, rather than an interval symmetrized about $r_1$ (i.e., we call particle ``1'' the one of two particles in the interaction at smaller radius), so particle pairs are manifestly not counted doubly.}  Inserting  this coarse-grained simulation phase-space density into Eq.~(\ref{eqn:prob}), the probability that particle 1 interacts in $\Delta t$ then becomes
\begin{equation}
    P_1(r_1,v_1,c_{\theta_1}) = \frac{\Delta t}{2} \left(\frac{m_p}{4\pi r_1^2 \Delta r}\right) \sum_{r_i=r_1}^{r_1+\Delta r} \frac{\sigma_{\rm tot}(v_{{\rm rel},i})}{m} v_{{\rm rel},i},
    \label{eq:prob2}
\end{equation}
where the sum is over all particles $i$ with radii $r_i$ in the interval $r_1 \to r_1 + \Delta r$.  The term in parentheses is the contribution of a single simulation particle to the density.

If the width $\Delta r$ is too large, the coarse-graining will smooth over important features of the phase-space density. If $\Delta r$ is too small, the number of particles in the sum will be small and random noise can affect the result. We choose $\Delta r=r_{i_1+j}-r_{i_1}$ to be the distance between particle $1$ (with index $i_1$---i.e., the $i_1$th particle from the center) and the $(i_1+j)$-th particle, where $j$ is some small fixed integer. We have taken $j=10$ in our default code and discuss below the dependence of the results on this assumption.

If some random number between 0 and 1 falls below $P_1$, then particle 1 scatters. If $j>1$,  we must then determine with which of the $j$ particles particle 1 interacts.  This is done by randomly choosing from them with probabilities weighted by their different values of $\sigma_{\rm tot}(v_{{\rm rel},j})v_{{\rm rel},j}$.  Of course, the validity of this simulation requires that $P<1$ and becomes better as $P$ becomes smaller.  Thus, a smaller time step will be required to resolve the dynamics with larger cross sections and densities.

Once the algorithm determines that two particles with velocities $\bfv_{1}$ and $\bfv_{2}$ scatter, the kinematics must be calculated. The center-of-mass (COM) velocity is $\bfv_{\rm cm} = (\bfv_1+\bfv_2)/2$ (for identical particles which we consider in this study) which can be used to boost in and out of the COM reference frame. The direction of the initial relative velocity (which is frame independent) is $\hat{v}_{{\rm rel},2}$.  The outgoing COM velocities, $\bfv_{{\rm cm}1}$ and $\bfv_{{\rm cm}2}$, each have magnitudes equal to $v_{{\rm rel},2}/2$ and are oriented in opposite directions to each other, with angles $\alpha$ with respect to $\hat{v}_{{\rm rel},2}$. This angle is chosen from the angular distribution given by $d\sigma/d\alpha$.  If the scattering is isotropic, this implies that $c_\alpha$ is chosen from a uniform distribution in the range $-1 \to 1$. Additionally, a polar angle $\phi_f$ around $\hat{v}_{{\rm rel},2}$ is chosen from a uniform distribution in the range $0 \to 2 \pi$. These angles fully define the outgoing COM velocities of both particles, and these are then boosted back into the halo frame and labeled $\mathbf{v}_{{\rm fin},k}$ (with $k=1$ or $2$). Each of these final velocities are then projected onto the $\hat{z}$ direction (the radial direction in the halo frame) and $c_{\theta,k} = (\mathbf{v}_{{\rm fin},k}\cdot\hat{z})/|\mathbf{v}_{{\rm fin},k}|$ which is then saved for the next time step.

\begin{figure}
\includegraphics[width=\columnwidth]{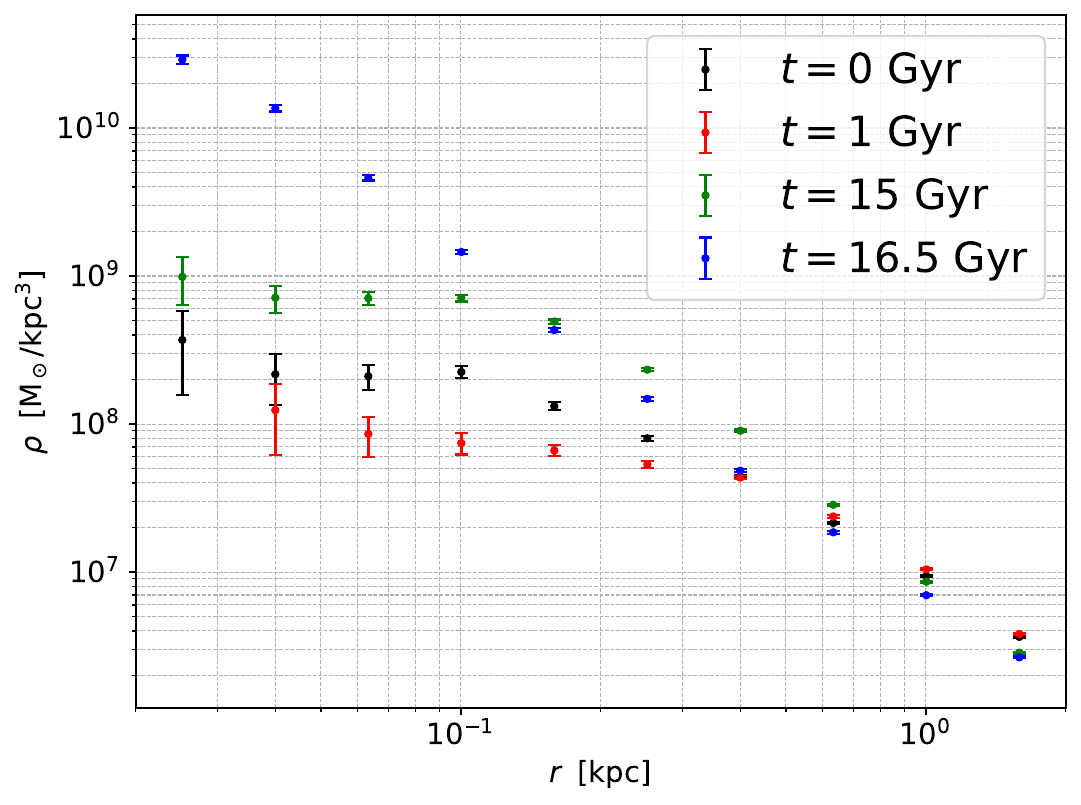}
     \caption{The time evolution of the inner density profile for a $1.15\times 10^{9}\,M_\odot$ NFW halo with scale radius $r_s=1.18$ kpc and density $\rho_s=2.73\times10^7\, M_\odot~{\rm cm}^{-3}$ and a velocity independent SIDM cross section (per unit mass) of 50 cm$^2$~g$^{-1}$ in a simulation with $10^5$ particles.  The simulation had  $1.25\times 10^7$  time steps with a 19.6 Gyr run time.  As seen, the initial profile (black) has an $r^{-1}$ power law at $r\lesssim 1$ kpc, which then very quickly flattens out by $t=1$ Gyr (red). The  steepening of the density at small radii expected from the gravothermal instability is seen at $t\simeq 15$ Gyr (green).  The $t=16.5$ Gyr points (blue) shows further steepening before the finite timestep ceases to resolve the effects of self-interactions.}
\label{fig:densityevolution}
\end{figure}

{\it Implementation and illustrative results.}  As a proof of principle, we have implemented this procedure into the recently developed {\tt NSphere} code \cite{Kamionkowski:2025uae}, and we have made this specific revision available at {\tt github.com/NSphere-SIDM/NSphere-SIDM}.  The self-interactions slow the code relative to the collisionless case, but for parameter values we have tried so far, it is only by tens of percent.

\begin{figure}
\includegraphics[width=\columnwidth]{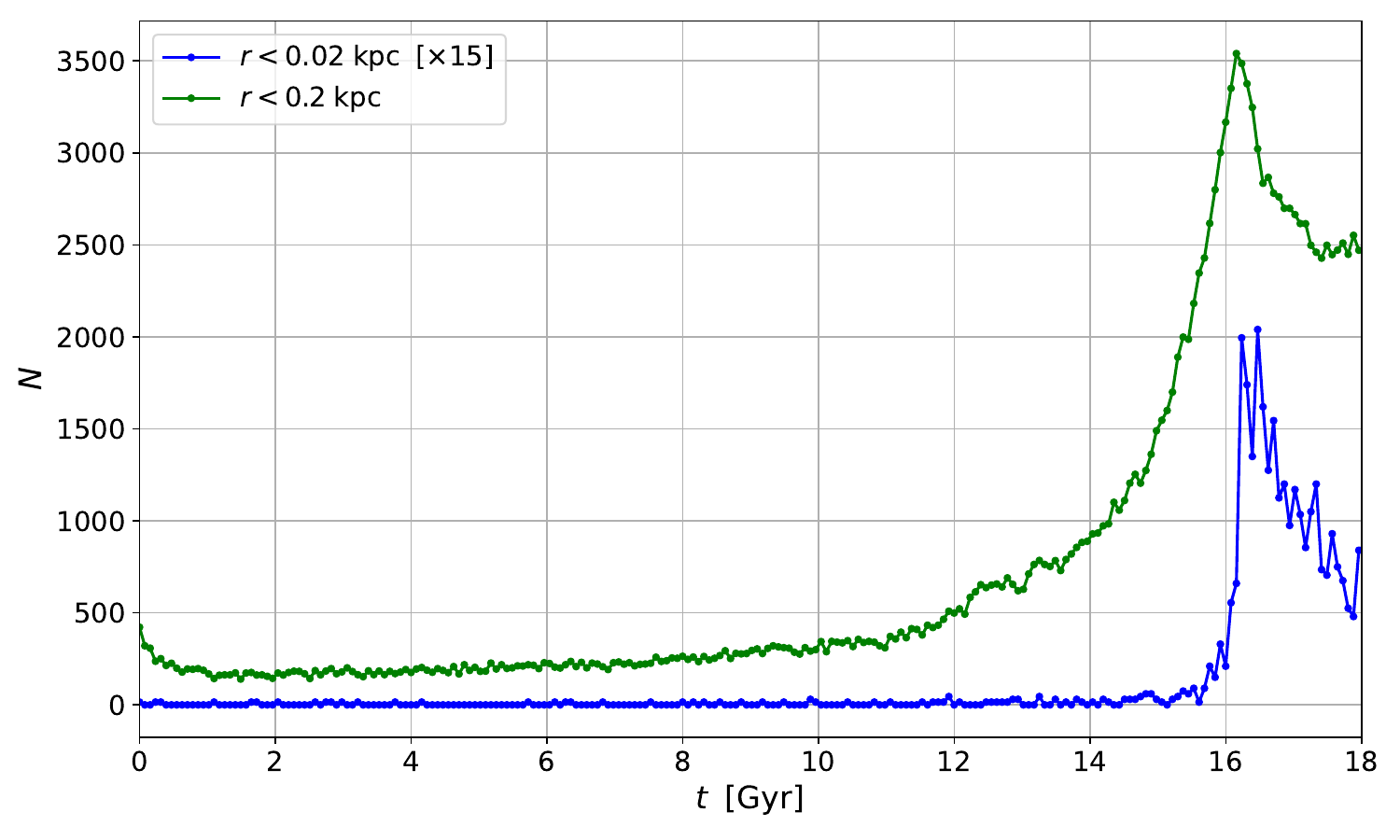}
     \caption{Time evolution of the number of particles in the innermost 0.02 kpc (blue) and 0.2 kpc (green) of the simulation shown in Fig.~\protect\ref{fig:centraldensity}.  The number within 0.02 kpc is scaled up by 15 so that the two curves have similar amplitudes.   The number of particles quickly decreases from its initial value (of a few hundred) and then slowly increases until about \mbox{$t=16$ Gyr.}, with a sharper increase in the innermost regions.   The decrease at times later than 16.5 Gyr is a numerical artifact resulting from the finite time resolution.}
\label{fig:centraldensity}
\end{figure}

To illustrate, we show results of a simulation of a halo with parameters that match those of one of the simulations in Ref.~\cite{Palubski:2024ibb}. I.e., we use an NFW halo with scale density $\rho_s = 2.73 \times 10^7\, M_\odot~{\rm kpc}^{-3}$ and scale radius $r_s=1.18$ kpc.  We use an SIDM cross section (per unit mass) $\sigma_{\rm tot}/m=50\,{\rm cm}^2~{\rm g}^{-1}$.  We choose the cross section to be isotropic and velocity-independent to compare with earlier work.  However, the simulation can be re-run with some other angular dependence by changing a single line in the code, and nontrivial velocity dependence can be added by adding another line of code.   We truncate the NFW halo near the virial radius $r_{\rm vir}=c r_s$ by multiplying the NFW halo by $\left[1+(r/r_{\rm vir})^{10} \right]^{-1}$ with concentration parameter $c=19$ (which leads to a total mass $M_h=1.15\times 10^9\, M_\odot$).  With these parameters, the dynamical time is $t_{\rm dyn}\equiv(G M_h/r_s^3)^{-1/2} \simeq 17.8$ Myr. 

The simulation models the effects of scattering accurately only if the timestep size is small compared with the scattering time $t_{\rm scat} = \left[\rho (\sigma/m)v \right]^{-1}$.  For our simulation, this evaluates, using the circular speed at $r_s$, to $t_{\rm scat} \simeq 110$ Myr, at a density $\rho_s$, and it scales with density roughly as $\rho^{-3/2}$.  Prior N-body codes use an adaptive time-step size to follow the evolution of high-density regions, and Ref.~\cite{Palubski:2024ibb} suggests that the time-step size should be $\kappa=0.002$ times the scattering time for precise results, with $\kappa=0.02$ giving results that differ by ${\cal O}(10\%)$.  Qualitatively, one expects the evolution to slow with larger timestep sizes, as the scattering rate is then underestimated.

Our current code is an augmentation of a code previously developed for more general problems involving spherically symmetric self-gravitating systems.  It therefore has a fixed time-step size.  In the simulations shown here we have chosen to have a timestep size of 0.00178 Myr, which should allow us to resolve densities $\rho \lesssim 10^{10}\,M_\odot\,{\rm kpc}^{-3}$ and  be good to ${\cal O}(10\%)$ only at $\rho\lesssim 10^8\,M_\odot\,{\rm kpc}^{-3}$.  This calculation is thus {\it very} inefficient, since only a small fraction of particles reach high densities, and those only for a short amount of time.  Even so, the simulation shown here was run in $\sim4$ hours on a 2023 MacBook Pro with a 12-core M2 processor.  We surmise that is should be straightforward to significantly speed up the code, and increase precision and accuracy, by incorporating an adaptive timesteping.

\begin{figure}
\includegraphics[width=\columnwidth]{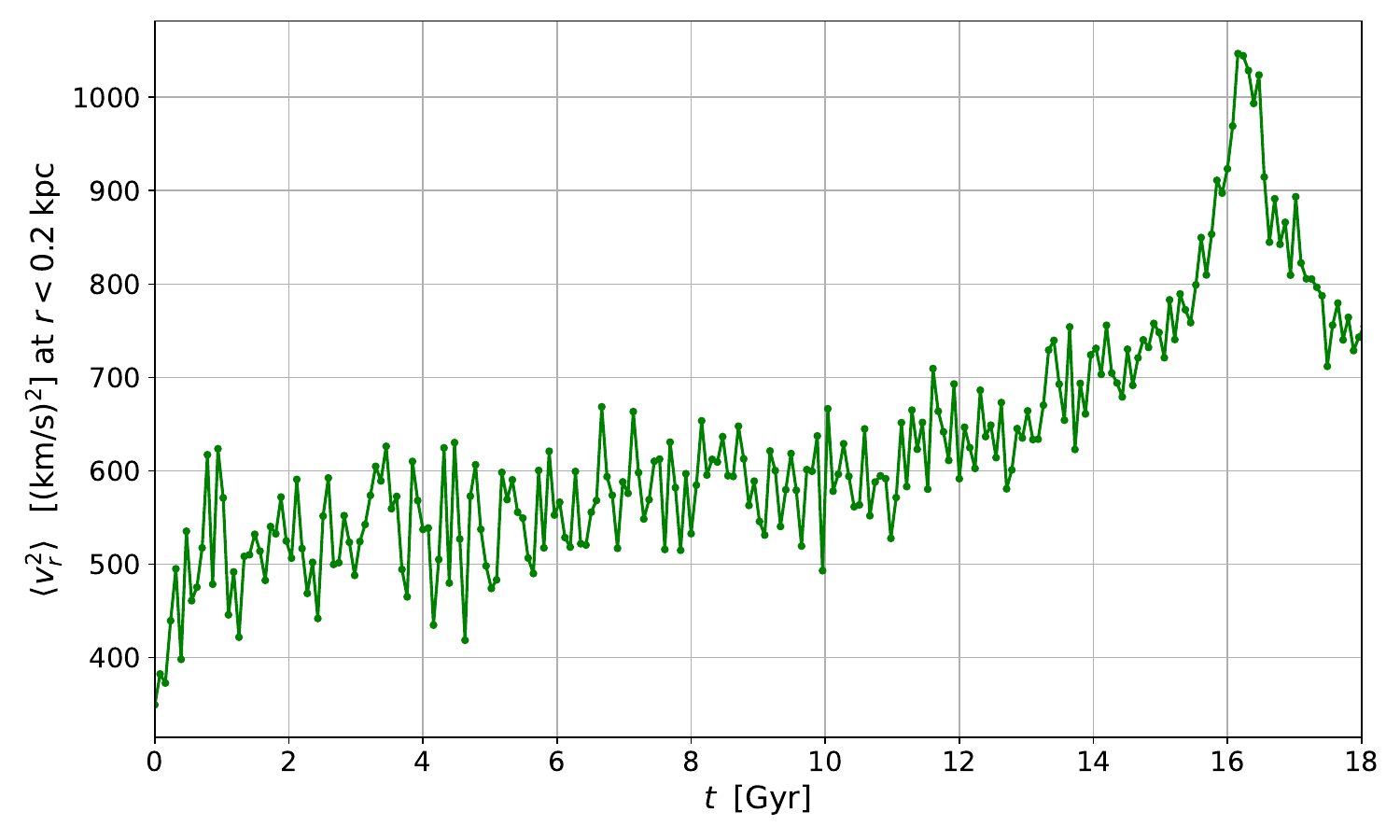}
     \caption{Time evolution of the radial velocity dispersion in the inner 0.2 kpc. In the first Gyr the initially cold population in the NFW profile is rapidly heated towards isothermality, and the dispersion then diverges as the core forms 16 Gyr later.}
\label{fig:dispersion}
\end{figure}

Fig.~\ref{fig:densityevolution} shows that the evolution of the density profile is consistent with what is expected.  The initial $r^{-1}$ density profile at $r\lesssim1$ kpc is smoothed out fairly quickly.  As seen in Figs.~\ref{fig:centraldensity} and \ref{fig:dispersion}, the central density and velocity dispersion in the inner regions then increase slowly for a while, but then increase rapidly approximately 16.5 Gyr later.  The blue points in Fig.~1 show the expected steepening into a dense core in the inner 0.01 kpc.  Subsequent time steps (not shown in Fig.~\ref{fig:densityevolution}) show an (artificial) decrease in the core density, as expected from the breakdown of the calculation when the density exceeds the critical value $10^{10}\,M_\odot\,{\rm kpc}^{-3}$ for the chosen parameters.

{\it Concluding remarks.}  Here we have shown that numerical evolution of self-gravitating spherical systems of self-interacting dark matter can be evolved numerically with an approach that is far more computationally efficient than---and sidesteps several technical challenges of---the N-body simulations traditionally used to tackle this problem.

We leave further detailed convergence tests for later specific science studies.  However, another simulation with a time-step size increased by a factor of 10 showed a longer collapse timescale (closer to 18 Gyr) consistent with our expectation that a larger timestep size leads to an (artificially) slower evolution. A simulation with that timestep size but 10 times more particles showed similar evolution.  We also see some run-to-run variance in the collapse time, consistent with sample variance---even with 100,000 simulation particles, there are no more than a few hundred particles in the inner 0.02 kpc. As discussed above, the scattering kernel was evaluated with $j=10$. However, similar results are obtained for $j=20$, with roughly the same computational time. However, a run with $j=2$ yielded results that were significantly different.

Several issues that arise in the traditional N-body approach are circumvented or alleviated with this new approach.  One does not need to consider finding the center of mass.  Furthermore, stochastic fluctuations are smaller for the same number of simulated particles given that they are denser in three phase-space dimensions than in six. In addition we need not implement force softening.  There are, as in the traditional approach, different ways of coarse-graining the phase-space distribution.  Whichever method is chosen, the scattering kernel is better sampled by coarse graining over a radius than it is over a three-dimensional volume and the mechanics of finding the nearest particles is simpler.  For precise results their dependence on the time-step size in this approach should be explored, as is the case for N-body simulations.  As seen in the simulations shown here, the time step can be taken to be fairly large for low-density regions, but a finer step size might be needed to resolve the high-density regions when the mean-free paths become very short.  We surmise that with a bit more effort, an adaptive time step can be incorporated, thus leading to an even more efficient code that can also follow the late-time evolution to even higher central densities.  Difficulties that arise in the N-body approach with energy non-conservation from multiple scattering on different CPUs are more easily dealt with in the new approach.

We have presented a new avenue to simulate the dynamics of spherical self-gravitating systems with self-interactions.  The approach capitalizes upon the reduction of the phase space in a spherically symmetric system from six dimensions to three.  A code that implements this approach allows for improvements in computational efficiency by orders of magnitude, allowing the types of simulations used in recent work to be carried out in a matter of minutes to hours on a laptop computer.  The approach is easily augmented to incorporate new phenomena, such as tidal stripping \cite{Nishikawa:2019lsc,Zeng:2021ldo}, accretion onto a central black hole \cite{Sabarish:2025hwb}, or to explore the effects of anisotropic velocity distributions.   It is simple to accommodate nontrivial angular and relative-velocity dependence of the scattering cross section, models with unequal masses, in inelastic \cite{Tucker-Smith:2001myb,Vogelsberger:2018bok} or dissipative scattering \cite{Krnjaic:2014xza,ONeil:2022szc}, or some combination of these.  It may also be employed to study the evolution of globular clusters by modeling hard gravitational scatters as self-interactions (although this will also require augmenting the system to account for multiple particle masses to model mass segregation).  We leave this, as well as further detailed exploration of specific SIDM models, to future work.

\medskip
We thank Andrew Robertson for useful discussions.  This work was supported at JHU by NSF Grant No.\ 2412361, NASA ATP Grant No.\ 80NSSC24K1226, the Guggenheim Foundation, and the Templeton Foundation.  This work was supported at UBC by funding from a NSERC of Canada Discovery Grant.   MK and OS thank the Center for Computational Astrophysics at the Flatiron Institute and the Institute for Advanced Study for hospitality.

\bibliography{refs}

\end{document}